\documentclass[12pt]{iopart}

\usepackage{graphicx}
\begin{document}
\letter{ Fermi Surface as a Driver for the Shape-Memory Effect in AuZn. }

\author{R.~D.~McDonald$^{\dag}$, J. C. Lashley$^{\dag}$, J. Singleton$^{\dag}$, P.A. Goddard$^{\dag}$, F. Drymiotis$^{\dag}$, N. Harrison$^{\dag}$, 
H. Harima$^{\ddag}$, M.-T. Suzuki$^{\ddag}$, A. Migliori$^{\dag}$
and J. L. Smith$^{\dag}$.}

\address{$^{\dag}$Los Alamos National Laboratory, Los Alamos, New Mexico 87545, USA}

\address{$^{\ddag}$ISIR Osaka University, Ibaraki, Osaka 567-0047 
and Department of Physics, Kobe University, 1-1 Rokko-dai Noda Kobe 657-8501, JAPAN}

\ead{rmcd@lanl.gov}

\begin{abstract}
Martensites are materials that undergo diffusionless, 
solid-state transitions. The martensitic transition 
yields properties that depend on the history of the 
material and may allow it to recover its previous 
shape after plastic deformation. This is known as 
the shape-memory effect (SME). We have succeeded 
in identifying the primary electronic mechanism 
responsible for the martensitic transition in 
the shape-memory alloy AuZn by using Fermi-surface 
measurements (de Haas-van Alphen oscillations) and 
band-structure calculations. This strongly suggests 
that electronic band structure is an important consideration 
in the design of future SME alloys.
\end{abstract}

\submitto{\JPCM} 

\maketitle

\begin{sloppypar}
The martensitic transition (MT) is widely encountered 
in nature, with examples being observed in cuprate 
superconductors \cite{Lavrov}, polymers \cite{Anseth}, 
transition metal alloys \cite{Bhatta} and 
actinides \cite{Lander}. A subclass of these materials 
exhibit the shape-memory effect (SME) where, following 
deformation of the low-temperature martensitic phase, 
they recover their high-temperature shape on warming through 
the transformation \cite{Schmer}.  One practical example 
of this phenomenon is the use of nitinol (NiTi-alloys) 
for stent implants used to increase flow in restricted 
blood vessels. Before insertion the stent is compressed 
and expands the vessel after warming to human body 
temperature \cite{Duerig}.

In elemental metals and simple metallic alloys, 
the band electrons act as the quantum-mechanical ``glue''
that binds the atoms in a particular structural 
arrangement~\cite{Abriko}; hence, these electrons 
are likely to be of great importance in spontaneous 
structural rearrangements such as the MT. Indeed, 
in several other classes of material, instabilities 
amongst the band electrons are already known 
to be responsible for structural phase transitions 
[see e.g., \cite{Lander},\cite{Gruner}]. 
In spite of this fact, and notwithstanding the 
importance of SME alloys in medicine and technology, 
very few opportunities have existed for experiments 
that elucidate directly the relationship between 
band electrons and the MT \cite{Ahlers}. 
To address this question, we have prepared a 
stoichiometric (disorder-free) SME alloy, AuZn, 
that shows a MT at 64 K, and made the first detailed 
measurements of its Fermi surface, the constant-energy 
surface in wave-vector space that defines the band-electron 
dynamics. By comparing the data with the first 
band-structure calculations of a martensite of 
this complexity (18 atoms per primitive unit cell 
with no inversion symmetry) we are able to ascertain 
the role of band electrons in the SME. This is an 
important step towards the design of future SME alloys 
with greater recoverable strain and improved functionality; 
because the high-temperature austentite, or parent, 
phase of AuZn shares the same B2 cubic symmetry as 
the more familiar NiTi \cite{Wayman} and AuCd \cite{Wechsler} 
SME alloys, it constitutes a paradigm for understanding 
the mechanism of this important effect.

Despite the similarities between AuZn and other SME alloys, 
the fact that its MT is only accessible using cryogens has 
made it much less studied. However, it is exactly this 
low-temperature transition that makes AuZn ideal for a 
study assessing the fundamental role of the electrons; 
in higher temperature martensites that role is often confused 
by the entropic contribution from the phonons. 

The transition takes place as follows \cite{Makita},\cite{Barsch}: 
as the temperature is lowered through the MT, 
the unit cell is distorted in the [110] shear direction. 
A strain and commensurate shuffle of every third unit cell 
results in a hexagonal primitive unit cell formed from 9 
primitive cubic cells of the parent phase. This structure 
can also be described in terms of its conventional 
rhombohedral unit cell; however, we limit our discussion 
to the primitive hexagonal unit cell as this is most 
informative when considering the relationship between the 
Brillouin zone  (the unit cell in wave-vector space) 
and a real-space distortion. 

Our single-crystal samples of AuZn were prepared by the 
method outlined in Reference \cite{Darling}. Resistivity 
data were acquired using the standard four-terminal technique 
at temperatures down to 500 mK in either a 33 T Bitter magnet 
or a 17 T superconducting magnet. Heat capacities were 
collected using a Quantum Design PPMS employing a thermal 
relaxation method. The excess heat capacity (i.e. that with the 
conventional phonon and electronic components subtracted) 
was computed in Reference \cite{Darling}. Magnetization 
measurements employed a capacitative torque magnetometer 
on a two-axis cryogenic goniometer, allowing samples to be 
rotated to all angles in the magnetic field without 
thermal cycling. Three planes of rotation intersecting 
symmetry directions of the parent phase were used, 
field sweeps being recorded every 3$\deg$. 
The initial sample orientation was determined by a Laue camera. 

\begin{figure}[tbp]
\includegraphics[height=8cm]{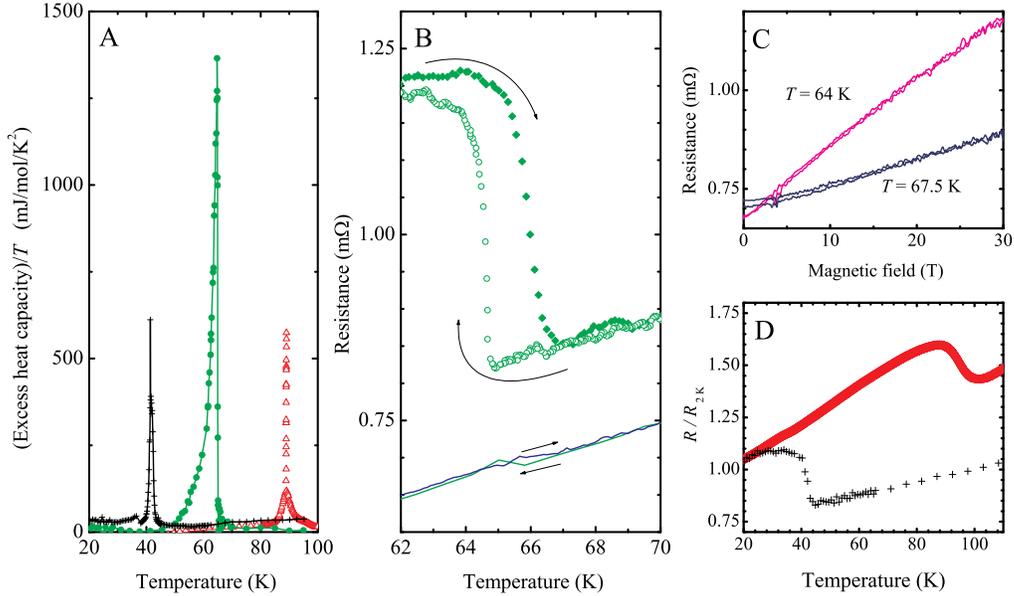}
\centering \caption{(A) Excess heat capacity [i.e. experimental 
heat capacity minus the slowly-varying background 
phonon contributions (11)] of Au$_{0.52}$Zn$_{0.48}$ (crosses), 
AuZn (filled circles) and Au$_{0.47}$Zn$_{0.53}$ (triangles). 
(B) Resistance of AuZn at zero magnetic field (curves) 
and at 25 T (points) versus temperature. 
In both cases, data for cooling and warming are shown. 
(C) Magnetoresistance of AuZn at 67.5 K (blue curves) 
and at 64 K (steeper red curves); data for both rising 
and falling field are shown. (D) Resistance 
(normalized to the value at 2 K) of Au$_{0.52}$Zn$_{0.48}$ 
(crosses) and Au$_{0.47}$Zn$_{0.53}$ (triangles) 
samples versus temperature (measured on cooling). 
Note how the zero-field resistance of the 
off-stoichiometric alloys exhibits a large step at the 
MT (D), whereas the resistivity of the equiatomic alloy 
(B) only develops a large step in magnetic field.} \label{AuZnFig1}
\end{figure}

Figure \ref{AuZnFig1}(A) shows the excess heat capacity 
for three single crystals of AuZn covering the range of 
composition exhibiting the SME. The transition from the 
high-temperature austenite to low-temperature martensite 
is clearly visible as a single peak in the excess heat 
capacity at temperatures 90 K (Au$_{0.47}$Zn$_{0.53}$), 
67 K (AuZn) and 40 K (Au$_{0.52}$Zn$_{0.48}$), confirming 
that this alloy has the lowest temperature shape-memory 
transition yet measured. Formation of martensite produces 
a vibrational entropy associated with the structural 
change along with a contribution from the number of 
distinct orientations that the variants can assume 
with respect to the parent phase \cite{Pops},\cite{Romero}. 
The absence of other features besides this peak is 
clear evidence that there are no precursor phases and 
that there is a simple atomic path through the transformation. 
Cryogenic deformation was used to demonstrate that the 
SME occurs in AuZn \cite{Darling}. 

A very significant change in magnetoresistance of 
AuZn at the MT is observed [Figs. \ref{AuZnFig1}(B), (C)]. 
In non-magnetic metals, magnetoresistance is caused by 
field-induced motion of electrons across the Fermi surface, 
resulting in an alteration of the time-dependent evolution 
of their velocities \cite{Abriko}. In the absence of magnetic order, 
the large change in magnetoresistance at the MT can only 
indicate one thing, a marked modification of the Fermi 
surface \cite{Abriko},\cite{Gruner}. The effect that any 
such modification has on the temperature-dependent 
resistivity can be quite small, depending as it does on 
the details of the density of states at the Fermi energy 
and the quasiparticle effective masses on either side of the 
transition. By contrast, the magnetoresistance is affected 
primarily by the shape of the electronic orbits and hence 
is a sensitive gauge of changes in Fermi-surface topology. 
This change in magnetoresistance, plus the large value 
(typically $\sim 30-40$) of the ratio of the sample resistance 
at 300 K to that at 4.2 K and the low temperature of 
the martensite transition suggest that a study 
of the Fermi-surface topology of AuZn would be informative, 
feasible and contain all of the essential physics. 
Hence the martensite phase was studied using the de Haas-van 
Alphen effect.

\begin{figure}[tbp]
\centering
\includegraphics[height=9cm]{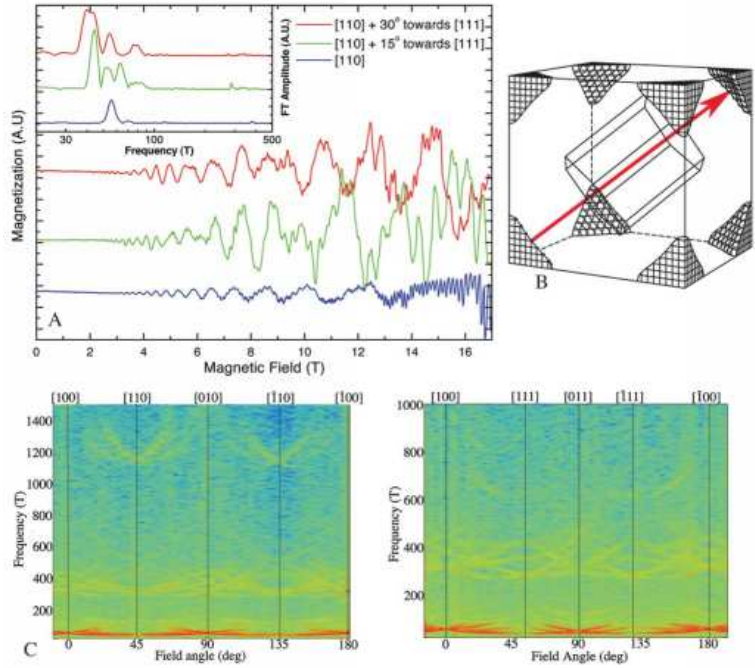}
\caption{(A) de Haas-van Alphen oscillations of a 
AuZn single crystal at a temperature of 0.75 K 
plotted as magnetization versus magnetic field for three 
different field angles. The inset contains Fourier 
transforms of the same data. (B) 
The hexagonal martensite-phase Brillouin zone 
enclosed within the cubic austenite Brillouin zone; 
the 14th band (hole) Fermi surface of the austenite 
is also shown, as is the proposed nesting vector (red arrow). 
The equivalence of the four possible orientations 
of the hexagon along the body diagonals of the cube 
is responsible for both the shape-memory effect 
associated with the transition and the cubic symmetry 
of the data in (C). (C) de Haas-van Alphen spectra plotted 
as Fourier amplitude (colour scale- red is high intensity, 
blue is background) versus frequency and field angle 
for rotations of the sample in $3^o$ steps about the 
austenite [001] direction (left) and [01-1] direction 
(right).} \label{AuZnFig2}
\end{figure}

In the de Haas-van Alphen effect, the frequencies, 
$F$ (in Tesla), of the magnetic quantum oscillations are 
proportional to extremal cross-sectional areas of the 
Fermi surface in the planes perpendicular to the magnetic 
field \cite{Abriko}. The oscillations can be directly 
related to the electronic density of states and Fermi-Dirac 
distribution function \cite{Abriko}. 
Typical raw de Haas-van Alphen data are shown in 
Fig. \ref{AuZnFig2}(A). The oscillations become visible 
at a field of around $B_{0}\sim 1.5$ T, 
suggesting that the electrons have a relatively 
long mean-free path, $\lambda$. Using the approximation 
$\lambda \sim (hF/ \pi e)^{1/2}/B_{0}$, where $h$ is Planck's 
constant and $e$ is the electronic charge \cite{Abriko}, 
we obtain $\lambda \sim 0.2 ~\mu$m. Such long mean-free paths, 
comparable to those in the best binary alloys \cite{Harrison}, 
imply that our Au$_{0.50}$Zn$_{0.50}$ samples are exceptionally 
free of impurities and defects. Therefore the SME is an intrinsic 
property of pure AuZn.

The inset to Fig. \ref{AuZnFig2}(A) shows Fourier transforms 
of typical data. As expected in a system with a 
large unit cell resulting from a complex shuffle, 
there is a plethora of low frequencies indicating 
a Fermi surface comprising many small pockets. 
To investigate further the Fermi-surface topology, 
magnetization data were recorded for many orientations of 
the sample in the field. Owing to the low symmetry of the 
martensite phase [Fig. \ref{AuZnFig2}(B)] three separate axes 
of rotation were chosen. 
Rotations of 180$\deg$ about each axis were used 
to allow the magnetic field to lie along a number of 
symmetry directions of the austenite phase. 
Figure \ref{AuZnFig2}(C) shows a mapping of the 
Fourier amplitudes of the various de Haas-van Alphen 
frequencies, each corresponding to a particular Fermi-surface 
cross-section. The frequencies show a very well-defined 
angular periodicity with several frequencies converging 
when the field is applied along the parent-phase symmetry directions. 
This observation demonstrates that there is a single-variant, 
low-temperature phase and that domains corresponding to 
all four possible orientations of the hexagonal distortion 
are present in roughly equal numbers, each possessing a 
well-defined angular relationship with each other and with 
the parent phase [Fig. \ref{AuZnFig2}(B)].
 The mean-free path of $\sim 0.2~\mu$m, inferred from 
the de Haas-van Alphen effect, represents an order-of-magnitude 
estimate for the linear dimensions of the domains.
 
In order to understand better the Fermi-surface topology, 
band-structure calculations were performed for the austenite 
and martensite phases. Atomic positions in the martensite 
phase were based on established group-theoretical 
arguments \cite{Barsch}. For the martensite phase the 
calculations are performed on the primitive hexagonal unit cell, 
containing 9 AuZn pairs \cite{Barsch}. The calculations include 
the effects of spin-orbit interactions by means of a full-potential 
linear augmented-plane-wave (FLAPW) method \cite{Harrison}. 
The exchange-correlation potential was modeled by means of a 
local-density approximation (LDA). 

\begin{figure}[tbp]
\centering
\includegraphics[height=9cm]{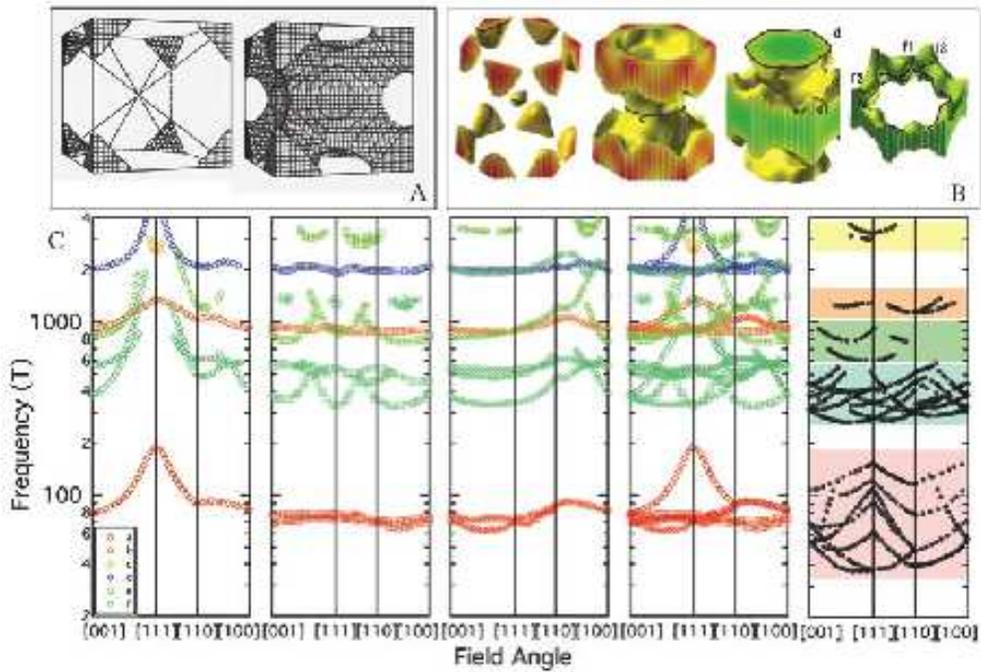} 
\caption{(A) The calculated Fermi surfaces 
for the austenite phase of equiatomic AuZn [left: 14th band, 
right: 15th band]. (B) The calculated Fermi surfaces 
for the martensite phase; the small letters label 
the Fermi surface orbits that give rise to various 
de Haas-van Alphen oscillations. (B) The three left-hand 
figures show de Haas-van Alphen frequencies as a 
function of field angle (with respect to the Austenite-phase 
axes) predicted by the band-structure calculations 
for equiatomic AuZn in the martensite phase. From left 
to right, the plots are for the martensite domain 
with the c-axis oriented along the austenite [111] direction,  
those along the [1-11]  and [-111] directions and 
that along the [11-1] direction respectively. The fourth 
plot is the sum of all domain orientations. The right 
hand plot shows the observed de Haas-van Alphen 
frequencies as a function of field angle (again 
with respect to the austenite-phase axes). The coloured 
bands are a guide to the eye, linking the observed frequencies 
with those predicted. The labels a-f indicate the Fermi-surface 
orbits responsible for each frequency [see (B)]. } \label{AuZnFig3}
\end{figure}

Figure \ref{AuZnFig3} shows the calculated Fermi-surface 
sections of the martensite phase [Fig. \ref{AuZnFig3}(B)] 
along with their predicted de Haas-van Alphen frequencies 
as a function of field orientation [Fig. \ref{AuZnFig3}(C)]. 
Note that three separate field-orientation dependences are 
shown to account for the four possible orientations of 
the martensite distortion. Theoretical frequencies from 
these orientations are compared with those observed in 
the experiments on the far right-hand side of Figure \ref{AuZnFig3}(C). 
The observed frequencies fall into the ranges encompassed 
by the predictions, and many of their angular dependences 
are reproduced qualitatively.
 
The success of the band-structure calculations enables us to 
suggest the way in which the Fermi-surface modification is 
associated with the MT. In the Austenite phase, two bands 
cross the Fermi energy, producing two Fermi-surface sheets 
[Figs. \ref{AuZnFig2}(B), \ref{AuZnFig3}(A)]. 
The sheet originating from the the 14th band has approximately 
planar faces orthogonal to the cubic <111> directions. 
Opposed planar faces of this type make a metal prone to a 
spontaneous distortion \cite{Abriko},\cite{Gruner}. If a single 
`nesting vector' [shown as a red arrow in Fig. \ref{AuZnFig2}(B) 
can map a large area of Fermi surface onto its 
symmetry-related equivalent, the electronic system 
can reduce its energy by opening a gap across those Fermi-surface 
sheets \cite{Gruner}. In real space, the nesting reveals 
itself as a spatial modulation of the electronic density 
and may be accompanied by a periodic distortion of the 
lattice via electron-phonon coupling. Such transitions 
occur provided the elastic energy cost of distorting 
the crystal to encompass the new periodicity, 
represented by the nesting vector, is outweighed by 
the energy saved by gapping the electronic system \cite{Gruner}. 
Systems with purely one-dimensional Fermi surfaces always 
exhibit this instability, and the resulting distortion 
is known as the Peierls distortion \cite{Abriko},\cite{Gruner}. 
In the case of AuZn's austenite phase, there are 
four equivalent proposed nesting vectors corresponding 
to the four body diagonals of the cubic Brillouin zone. 
This explains the four crystallographically-related 
domains found in the martensite phase [\ref{AuZnFig2}(B)].

In a low-symmetry metal with a one-dimensional Fermi surface, 
a Peierls distortion can be either commensurate with the 
lattice if the Fermi momentum is a rational fraction 
of the momentum at the Brillouin-zone boundary, or incommensurate 
if it is not \cite{Gruner}. In either case, to first order, 
the lattice distortion moves successive planes alternately 
closer together and further apart, thus introducing a new 
periodicity without changing the sample length. 
In a high-symmetry metal with multiple planar Fermi-surface 
sheets the situation may be different. If the separation 
of the planar sections of Fermi surface is not a rational 
fraction of a dimension of the Brillouin zone, a 
crystallographic distortion resulting in a net 
length change along the preferred direction can change 
the relative Fermi- and Brillouin-zone momenta 
(this cannot occur in a one-dimensional metal), 
facilitating a commensurate distortion that gaps the 
Fermi-surface sheet in question. A commensurate 
distortion is preferred because the reduction in 
electronic energy is greater than for the incommensurate case. 
This is precisely the situation that measurements indicate 
for the MT in AuZn, thus providing an explanation for the SME: 
in AuZn, the momentum of the 14th band Fermi surface is 
close to, but not exactly, 2/3rds that of the Brillouin-zone boundary 
such that the distortion must not only triple the unit cell 
to gap these Fermi-surface sections but also change the 
length of the sample in that direction so as make the 
distortion commensurate. It is the net length change 
of one of the sample dimensions relative to the others 
that gives AuZn its SME properties, enabling it to strain in 
response to an applied stress by changing the relative 
number of domains oriented in each direction. 
The strain is hence recoverable upon warming through 
the MT, and AuZn exhibits the SME. The experiments and 
calculations reported here have not only brought 
this mechanism to light, but strongly suggest important 
considerations for the design of future SME alloys, 
{\it i.e.}, the ideal requirements for the austenite phase 
are near-planar Fermi-surface sections linked by 
geometrically-degenerate nesting vectors that are close 
to commensurate with the Brillouin zone.
 
This electronic mechanism may also explain the trend in 
MT temperature with composition [\ref{AuZnFig1}(A)]. 
Zinc contributes two electrons to the filling of the 
upper bands in AuZn, whereas gold only contributes one. 
As the band filling increases with zinc composition, 
the size of the hole Fermi-surface pockets in the austenite 
phase will shrink slightly. More importantly, they 
will `flatten out', so that a distortion will 
be able to gap a greater fraction of states at 
the Fermi energy. This will increase the amount of 
free energy gained and raise the transition temperature. 

In summary, magnetoresistance data show that the 
martensitic transformation in AuZn is associated with a 
single, distinct rearrangement of the Fermi surface. 
In combination with heat-capacity experiments, this 
demonstrates that the transformation is free of the 
complications of precursor phases. Furthermore, the 
low transition temperatures imply that effects due to 
diffusion \cite{Gupta} and vibrational entropy can be ruled out. 
The de Haas-van Alphen data suggest an order-of-magnitude 
length scale for the martensite domains and are in agreement 
with band structure calculations. Most importantly the 
data and calculations provide direct evidence 
about the role of the band-electron system and its 
Fermi surface in the SME. This indicates that 
band-structure/property relations are an important 
consideration for the design of future SME alloys.

This work is supported by the U.S. Department of 
Energy (DOE) under Grant No. LDRD-DR 20030084. 
Part of this work was carried out at the 
National High Magnetic Field Laboratory, which is 
supported by the National Science Foundation, the 
State of Florida and DOE. We thank M. Ahlers, 
T. Lookman, and D. Lieberman for many informative sessions. 
We also thank Luis Balicas and Ali Bangura for experimental assistance.

\end{sloppypar}

\end{document}